\documentclass[traditabstract]{aa}

\usepackage[authoryear]{natbib}
\bibpunct{(}{)}{;}{a}{}{,} 

\usepackage{graphicx}
\usepackage[breaklinks,colorlinks=true,citecolor=black,linkcolor=black,urlcolor=blue,bookmarks=true]{hyperref}
\usepackage{txfonts}
\usepackage[utf8]{inputenc}
\usepackage{marginnote}
\usepackage{color}
\begin{document}

\title{{\huge{\it Research Note}}\\
\mbox{6-meter} telescope observations of three dwarf spheroidal galaxies with very low surface brightness}
\titlerunning{\mbox{6-meter} telescope observations of dSphs}

\author{
D.I.\ Makarov$^{1,2}$\fnmsep\thanks{\email{dim@sao.ru}}
\and
M.E.\ Sharina$^{2}$
\and
V.E.\ Karachentseva$^{3}$
\and
I.D.\ Karachentsev$^{2}$
}
\authorrunning{D.I.\ Makarov et al.}

\institute{ Universit\'e Lyon~1, Villeurbanne, F-69622, France;
CRAL, Observatoire de Lyon, St. Genis Laval, F-69561, France; CNRS
UMR 5574, France \and Special Astrophysical Observatory of the
Russian Academy of Sciences, Nizhny Arkhyz, Karachai-Cherkessia
369167, Russia \and Main Astronomical Observatory, National
Academy of Sciences of Ukraine, 27 Akademika Zabolotnoho St.,
03680 Kyiv, Ukraine }

\date{Received date / accepted date}


\label{firstpage}

\abstract {Dwarf spheroidal galaxies
(dSphs) are mostly investigated in the Local Group. DSphs are
difficult targets for observations because of their small size and
very low surface brightness. Here we measure spectroscopic and
photometric parameters of three candidates for isolated dSphs,
KKH\,65$=$BTS23, KK\,180, and KK\,227, outside the Local Group.
The galaxies are found to be of low metallicity and low velocity
dispersion. They are among the lowest surface brightness objects
in the Local Universe. According to the measured radial
velocities, metallicities, and structural and photometric parameters,
KKH\,65 and KK\,227 are representatives of the ultra-diffuse
quenched galaxies. KKH\,65 and KK\,227 belong to the outer parts of the groups
NGC3414 and NGC5371, respectively. KK\,180 is located in the Virgo
cluster infall region.}

\keywords{
galaxies: dwarf --
galaxies: distances and redshifts --  galaxies: fundamental parameters --  galaxies: photometry
}
\maketitle

\section{Introduction}
\label{sec:intro} \citet{2010Ap.....53..462K} selected nine
candidates for isolated  dSphs in the
Local Supercluster. Here we study three galaxies
(Table~\ref{tab:allres}) from this list.
KKH\,65$=$BTS\,23 was discovered by \citet{1990AA...228...42B} and
rediscovered in radio observations by \citet{2001AA...366..428K}.
KK\,180 and KK\,227 were first found by
\citet{1998AAS..127..409K}.

As distinct from dwarf irregulars, dSphs tend to be close
companions of bright giant galaxies. 
Many theoretical models explain the formation of early-type dwarf galaxies
 by interaction of their progenitors 
with massive neighbors and subsequent loss of gas (e.g., \citet{1974Natur.252..111E}; \citet{2012MNRAS.419..971D}
and references therein).
Isolated early-type dwarf galaxies are extremely rare within 10 Mpc \citep{2001AA...379..407K,
2012MNRAS.425..709M, 2015MNRAS.447L..85K}. 
Their formation mechanism is still poorly understood.  
Their observational properties are of particular interest.
Isolated dSphs were not
able to interact with massive neighbors during their lifetimes.
Consequently, their existence is evidence in favor of different gas-loss factors, for example, interaction with other small galaxies, gaseous filaments in the intergalactic medium
\citep{2013ApJ...763L..41B}, or cooling and feedback processes in
the early Universe during the epoch of reionization
\citep{2009ApJ...693.1859B}. N-body cosmological simulations
\citep[e.g.,][]{1999ApJ...522...82K,2005ApJ...629..259R,2014arXiv1405.4523K}
predict $ \sim30$ times more massive isolated dwarf galaxies than
are actually observed. dSphs with  very low surface brightness (LSB) are good candidates for this
missing population because they are barely detectable in the
optical and radio bands.

In this paper, we report the results of extensive spectroscopic
observations carried out with the Russian 6 m telescope. The
observed data allowed us to identify neighbors of these three
galaxies and estimate group membership distances and physical
parameters for them: sizes, masses, and luminosities.

\begin{table}[h]
\caption{\label{tab:allres} Main data for the galaxies: (1-2)
equatorial coordinates for the epoch J2000.0; (3-6) color excess
due to Galactic extinction, galaxy size at the isophote
$\textrm{SB}_V \sim 28$ mag~arcsec$^{-2}$, absolute V magnitude
and effective radius from our photometry (Sect.~\ref{surfphot});
group membership distance in Mpc, (8-10) summary of the data
acquired from the spectroscopic analysis (Sect.~\ref{sec:specres}):
heliocentric radial velocity and its dispersion, and metallicity;
(11) radial velocity with respect to the center of the LG calculated according to  \citep{1996AJ....111..794K}.
The superscripts refer to (1) \citep{2001ApJ...546..681T} , (2) \citep{2000ApJ...529..745F}, 
and (3) \citep{2008AJ....135.1488T}.}
\begin{tabular}{lccc}
\hline\hline
Data/Object                 & KKH\,65     & KK\,180     & KK\,227   \\
\hline
RA(2000.0)                  &  10 51 59.2 & 13 04 30.2  & 13 56 10.1 \\
DEC(2000.0)                 & +28 21 45   & +17 45 32   & +40 18 12 \\
$ E(B-V)$                   & 0.02        & 0.02        & 0.01       \\
Diameter~($ \arcmin$)       & 0.94        & 1.5         & 0.82        \\
$M_V$ (mag)                 &  -15.08     &  -14.98     &  -15.22     \\
$R_{e,V}$ (kpc)             &   2.40      &  1.62       &  2.75       \\
\smallskip
$D^{\rm GrMem}_{\sun}$~(Mpc) & 25.2$^1$    & 16.4$^2$    & 29$^3$   \\
$V_h$~(km~s$^{-1}$)         & 1350$\pm$50 & 687$\pm$40  & 2125$\pm$65 \\
$\sigma_V$~(km~s$^{-1}$)    & 26:          & 18:          &  17:        \\
$\textrm{[Fe/H]}$~(dex)     & $-1.5\pm0.3$      & $-1.65\pm0.52$  & $-0.9\pm0.4$     \\
$V_{\rm LG}$~(km~s$^{-1}$)  & 1301$\pm$50 & 609$\pm$40  & 2186$\pm$65 \\
\hline\hline
\end{tabular}
\end{table}

\section{Observations and data reduction}
\label{sec:Obs} Spectroscopic observations were carried out in
2014 and 2015 with the \mbox{6-meter} telescope of the Special
Astrophysical Observatory of the Russian Academy of Sciences (SAO
RAS). The SCORPIO
spectrograph\footnote{http://www.sao.ru/hq/lon/SCORPIO/scorpio.html}
was mounted in the primary focus \citep{2005AstL...31..194A}.
Table~\ref{tab:log} includes the date, exposure time, seeing, and
heliocentric velocity correction for each galaxy.\begin{table}
\caption{\label{tab:log}Log of spectroscopic observations.}
\begin{tabular}{lclr}
\hline\hline
Object & Date       & Exposure & Seeing    \\ 
       & (y/m/d) & (s)      & ($\arcsec$)  \\ 
\hline
KKH\,65  & 14.02.01       & 4 x 1200, 1000 & 2.6     \\ 
       & 14.02.03       & 4 x 1200         & 1.5     \\ 
KK\,180  & 14.02.01       & 4 x 1200       & 2.6     \\ 
       & 14.02.03       & 2 x 1200, 600    & 1.5     \\ 
KK\,227  & 14.02.03       & 4 x 1200       & 1.5     \\ 
       & 15.02.15       & 5 x 1200         & 2.0     \\ 
       & 15.02.18       & 4 x 1200         & 1.4     \\ 
\hline\hline
\end{tabular}
\end{table}
\begin{figure}
\includegraphics[width=0.48\textwidth]{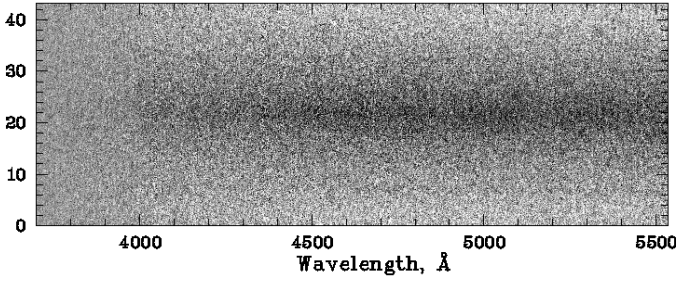}
\caption{\label{fig:2d} One-hour long-slit exposure of KK180, the
brightest galaxy of the sample. The vertical axis indicates the
position along the slit in arcseconds.}
\end{figure}
\begin{figure}
\includegraphics[width=0.32\textwidth,angle=-90]{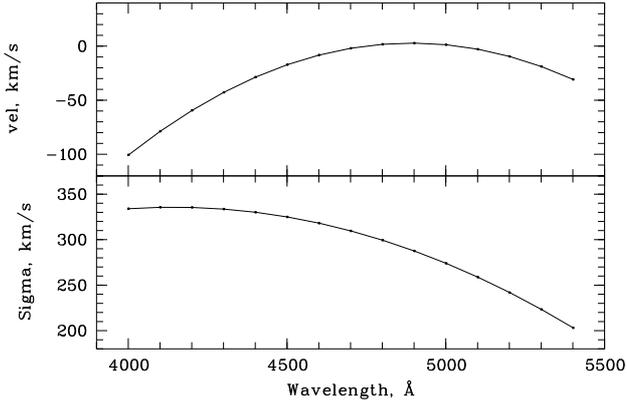}
\caption{\label{fig:lsf} Variation of the measured velocity and
instrumental velocity dispersion in the twilight spectrum as a function of  wavelength.}
\end{figure}
 \begin{figure}
\begin{tabular}{p{1.0\textwidth}}
\includegraphics[width=4.85cm,angle=-90,bb=76 55 510 850]{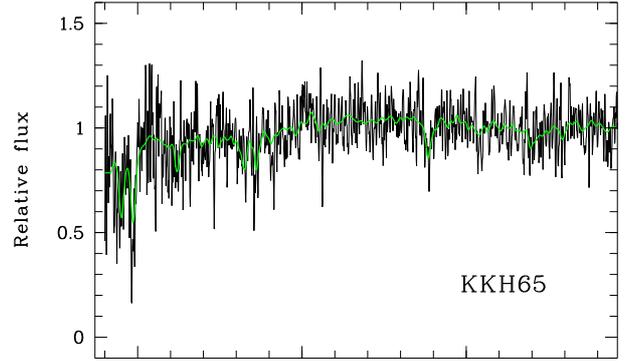} \\
\includegraphics[width=4.85cm,angle=-90,bb=76 55 510 850]{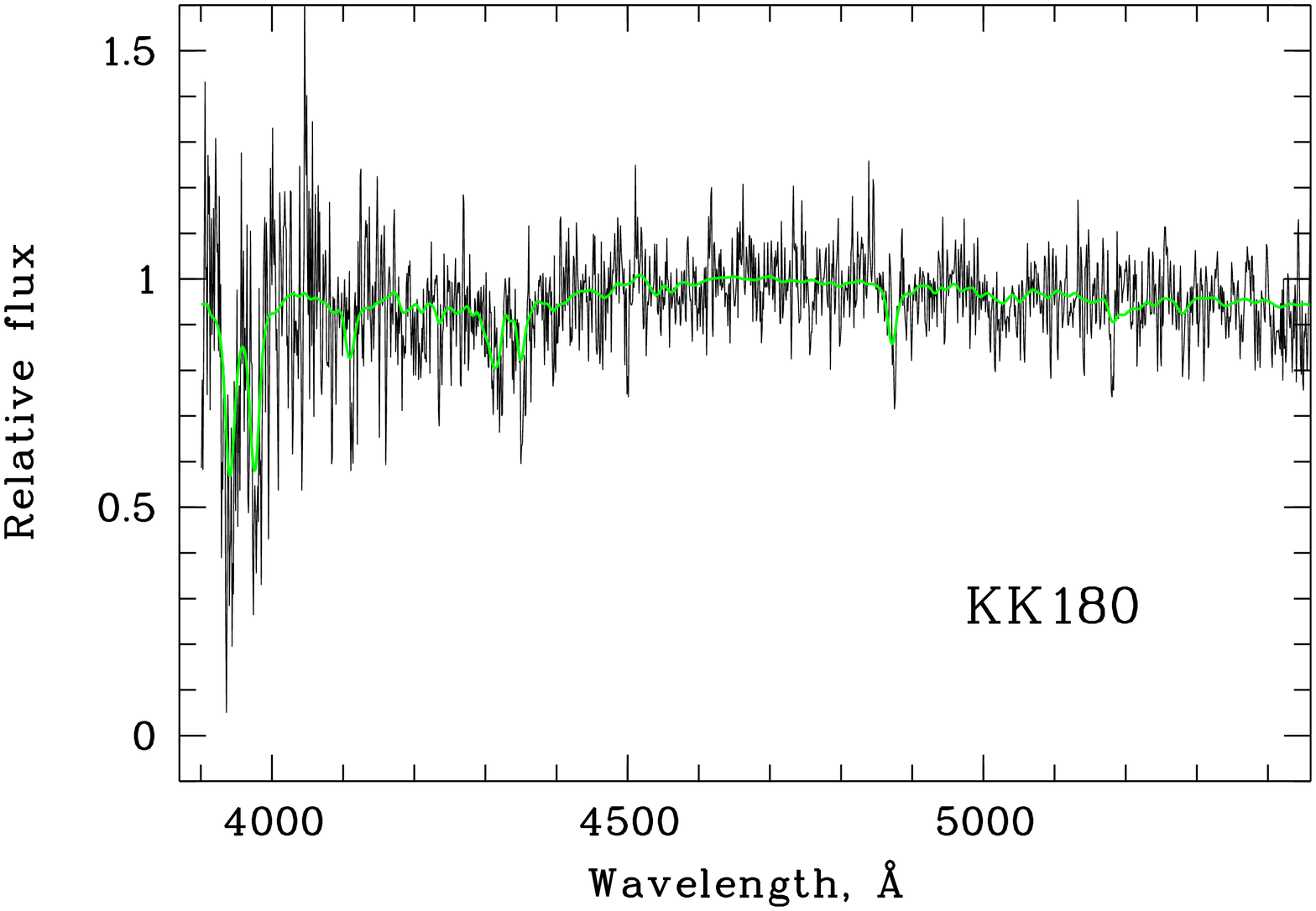} \\
\includegraphics[width=4.87cm,angle=-90,bb=76 55 510 850]{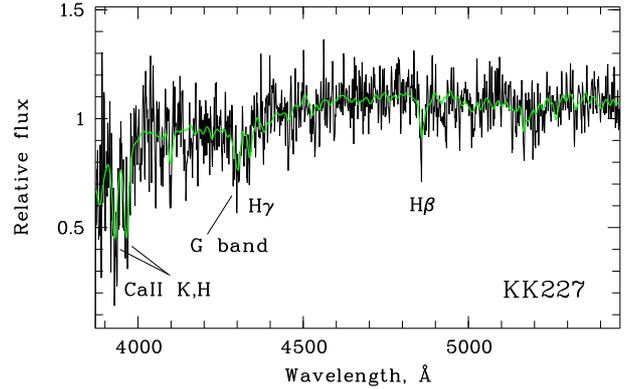} \\
\\
\\
\end{tabular}
\caption{\label{fig:KK180outulyss} Integrated spectra of the
stellar light of three dSphs (black) (top: KKH\,65, middle:
KK\,180 and bottom: KK\,227) in comparison with a composite model
(green). The fitting was carried out using the {\sc ULySS}
program, PEGASE-HR SSP model, the ELODIE stellar library, and the
LSF of the spectrograph.}
\end{figure}
The atmospheric transparency was very good during the nights.
Because of the extremely low brightness of the galaxies (see
Fig.~\ref{fig:2d}, Sect.~\ref{surfphot}), the total exposures for
each object were long. The long slit was always 6\arcmin $\times$
1\arcsec and the field of view was 6\arcmin$\times$6\arcmin. The
instrumental setup included the CCD detector EEV\,42-40 with a
pixel scale of $0.18$\arcmin~pixel$^{-1}$ and the grism VPHG1200B
with a resolution of FWHM $\sim5$~\AA{} and a reciprocal
dispersion of 0.9~\AA/pix in the spectral range of 3700--5500~\AA.
Every night, we obtained spectra of twilight, spectrophotometric
standard stars,
 and radial velocity standard stars.
The position angle of the slit was always nearly parallactic for
all the galaxies except KK\,227. Spectra reductions revealed the
existence of a faint distant object seen through the stellar body
of KK\,227 with a spectrum resembling that of a quasi-stellar
object (QSO) (see Appendix~\ref{sec_QSO}).
Therefore, this dSph was reobserved in 2015 with a different
position of the slit (Fig.~\ref{fig:QSO}) to improve the
signal-to-noise ratio in the one-dimensional total spectrum of
KK\,227.

The standard spectral data reduction was performed using the European Southern Observatory Munich Image Data
Analysis System ({\sc MIDAS}) \citep{1983Msngr..31...26B}
and the Image Reduction and Analysis Facility ({\sc IRAF}) software system
\citep{1993ASPC...52..173T}.
The accuracy of the wavelength calibration was $\sim0.16$~\AA.

\section{Spectroscopic results}
\label{sec:specres} To derive radial velocities of our
objects, we used the \textsc{ULySS} package
\citep{2008MNRAS.385.1998K,2009AA...501.1269K} with the PEGASE-HR
model grids \citep{2004AA...425..881L}, the Salpeter IMF
\citep{1955ApJ...121..161S}, and the ELODIE
\citep{2001AA...369.1048P} stellar library. To extract kinematic
information from the spectra, we first determined the line
spread function (LSF) of our spectrograph using the prescriptions
of \cite{2008MNRAS.385.1998K}.
The LSF was approximated by comparing a model spectrum of the Sun
with the twilight spectra taken during the same night. The result
of the comparison is shown in Fig.~\ref{fig:lsf}. It demonstrates the change with wavelength in the measured radial velocity and instrumental velocity
dispersion. These dependences were taken
into account when we computed radial velocities using
\textsc{ULySS}. Figure~\ref{fig:KK180outulyss} shows the results
of the spectral fitting for the galaxies.
One-dimensional total spectra of the galaxies have a quite low
signal-to-noise ratio (S/N) per pixel in the middle of the spectral
range: $S/N\sim20$ for KK\,180, $\sim10$ for KKH\,65, and $\sim8$
for KK\,227.
 Our spectra allowed us to derive radial velocities,
metallicities, and approximate velocity dispersions of KKH\,65,
KK\,180, and KK\,227 (Table~\ref{tab:allres}).

\section{Surface photometry results}
\label{surfphot} We performed surface photometry of the galaxies
using Sloan Digital Sky Survey (SDSS) images in the $g$, $r,$ and
$i$ bands and fitted their surface brightness profiles using the
Sersic function \citep{1968adga.book.....S}. The results are shown
in Figs.~{\ref{fig:serskkh65}-- \ref{fig:serskk227} and are summarized
in Tables~\ref{tab:allres} (rows 4-6), \ref{tab:photres}, and
\ref{tab_gri}. Table \ref{tab:photres} contains the following
columns: (1)~galaxy name,
(2)~apparent $V$ magnitude (mag) and mean color $(V-I)$,
(3)~colors $(B-V)$ and $(R-I)$, (mag), (4)~Sersic index $n$ of the
major-axis profile fit in the $V$ band, (5--7)~the corresponding
$V$-band central SB, $\mu_{0_s}$ in mag arcsec$^{-2}$), effective
radius (arcsec), effective SB in mag arcsec$^{-2}$ and apparent V
magnitude integrated within the effective radius. The photometric
data in Table~\ref{tab:photres} were corrected for Galactic
extinction.
 The procedure of the data analysis is described in Appendix~\ref{SecA1}}.
\begin{table}[h]
\caption{\label{tab:photres} Surface photometry and Sersic
function fitting results for the galaxies.}
\begin{tabular}{lrrrrrr}\hline\hline
Name &  $V  $& $B-V$ & $n_{V}$ & $\mu_{e,V}$      & $\mu_{V,0}$ & $r_{e,V}$ \\
      &$V-I$ &$R-I$  &           &    $V_e$           &               &             \\ \hline
KKH\,65 & 16.93&  0.61 & 0.92      & 25.97              &  24.31        & 19.8        \\
      & 0.93 &  0.57 & $\pm$0.04 & 17.55              &   $\pm$0.03   & $\pm$0.6    \\
KK\,180 & 16.09&  0.73 & 0.95      & 25.38              &  23.67        & 20.5        \\
      & 1.06 &  0.59 &  $\pm$0.02& 16.86              &  $\pm$0.02    &  $\pm$0.4   \\
KK\,227 & 17.09&  0.71 & 0.88      & 26.06              &  24.50        & 19.6        \\
      &  1.17&  0.66 & $\pm$0.05 &  17.73             &  $\pm$0.04    &  $\pm$0.8   \\
\hline\hline
\end{tabular}
\end{table}

We estimated the $K_s$ magnitudes of KKH\,65, KK\,180, and KK\,227
using the transformations between the SDSS and 2MASS photometric
systems \citep{2011MNRAS.417.2230B}: $Ks_0 =g - 1.907\cdot(g-r) -
1.654\cdot(r-i)-0.684 $. The $K_s$ luminosities were converted into
stellar masses of the galaxies using $M_{Ks_{\sun}} = 3.29$
\citep{2007AJ....133..734B} and $(M/L)_{Ks} = 0.9$ in solar units
\citep{2005MNRAS.362..799M}.

\section{Group membership distances of KKH\,65, KK\,180, and
KK\,227 and association with other galaxies} 
\label{sec:ii}
Identifying possible neighbors of the three dSphs was an important
step of the work. Understanding the real isolation status of the
galaxies and their physical properties requires information on
their redshift-independent distances. To perform this task, we
first searched for possible neighbors within a projected distance
of $\sim$500 kpc and with differences in radial velocities slower
than $\sim$500 km~s$^{-1}$. Next, we used the group-finding method
by \citet[hereafter MK11]{2011MNRAS.412.2498M}.
 The MK11 algorithm uses radial velocities of objects,
 projected distances between them, and their masses.
To apply this method, we used the data obtained in our study for
the three dSphs and the corresponding data for the surrounding
galaxies extracted from the literature using the Aladin sky atlas
\citep{2000AAS..143...33B} and the HyperLEDA database
\citep{HyperLEDA}. We used the value of the Hubble constant:
$H_0=73$~km~s$^{-1}$~Mpc$^{-1}$.
Appendix~\ref{association} contains the detailed results of group
member searches for our galaxies. Tables~\ref{tab:n3414gr}and
\ref{tab:kk180gr} list the data for the projected neighbors of
KKH\,65 and KK180.

\section{Discussion and conclusions}
The estimated metallicities and approximate velocity dispersions
(Table~\ref{tab:allres}) of KKH\,65, KK\,180, and KK\,227 are similar of dSphs in the LV
\citep[e.g.,][]{2013ApJ...765...38G}. In the following, we examine how typical the other derived parameters are.

{\bf KKH\,65.} The nearest bright galaxy to KKH\,65 in projection to the sky is the peculiar S0-galaxy NGC\,3414.
The mean radial velocity of the NGC\,3414 group with respect to the LG
is $V_{LG}=1298\pm117$~km~s$^{-1}$ (MK11).
This value agrees well with the velocity of KKH\,65
$V_{LG}=1301\pm50$~km~s$^{-1}$ found by us.
Thus, KKH\,65 is a probable member of the group.
We here adopted the distance to the NGC\,3414 group
measured by \citet{2001ApJ...546..681T} using
SB fluctuations: $D_{\sun}=25.2$~Mpc.
The corresponding projected separation between KKH\,65 and
NGC\,3414 is 188 kpc, the value typical for dSphs
 \citep{2005nfcd.conf..295K}.
At this distance, KKH\,65 is larger and more luminous
($M_V=-15.08$~mag, $R_{e,V}=2.4$~kpc) than dSph neighbors of
our Galaxy \citep[e.g.,][]{2012AJ....144....4M}. The luminous mass
of KKH\,65 calculated using $Ks_o=15.3$~mag (Sect.~\ref{surfphot})
is $9\times10^7 M_{\sun}$.

{\bf KK\,180.} The nearest bright galaxy to KKH\,65 in projection to the sky is the SBc galaxy UGC\,8036. 
Its radial velocity with respect to the LG,
$V_{LG}=844$~km~s$^{-1}$, is very close to the velocity of
KK\,180 $V_{LG}=609 \pm 40$~km~s$^{-1}$ that we found. A small
galaxy group around UGC\,8036 resides in the outskirts of the
Virgo cluster \citep[e.g.,][]{2008ApJ...676..184T}.
We here adopted the distance to KK\,180 equal to the distance to the Virgo cluster:
 $D_{\sun}=16.4$~Mpc \citep{2000ApJ...529..745F}.
Based on this, KK\,180 is located at a projected distance of
1.39Mpc from M87, that is, within the virial radius of the Virgo
cluster $R_v=1.8$~Mpc \citep{1980ApJ...242..861H}. Its luminosity,
effective radius, and mass are $M_V=-14.98$~mag,
$Ks_o=14.14$~mag, $R_{e,V}=1.6$~kpc, $Mass_{Ks}=1.1\times10^8
M_{\sun}$.
KK\,180 is as luminous as KKH\,65, but by 1.5 times more compact.

{\bf KK\,227}. This galaxy most likely belongs to the NGC\,5371 group
of 55 galaxies (MK11) according to its radial velocity and
position in the sky (Table~\ref{tab:allres}).
The distance to the Sbc galaxy NGC\,5371 was estimated using the
Tully-Fisher relation by \citet{2008AJ....135.1488T}:
$D_{\sun}=29$~Mpc.
The radial velocity of NGC\,5371 with respect to the center of the
LG is $V_{LG}=2640$~km~s$^{-1}$ \citep{2001AA...378..370V}. A mean
velocity of the group is $V_{LG}=2615$~km~s$^{-1}$ , and its
dispersion is $\sigma_{V}=195$~km~s$^{-1}$ (MK11). If we accept
the group membership distance $D_{\sun}\sim 29$ Mpc for KK\,227,
then the projected separation between KK\,227 and NGC\,5371 is
$\sim74$~kpc, $M_V=-15.22$~mag, $Ks_o=15.0$~mag,
$R_{e,V}=2.75$~kpc, and $Mass_{Ks}=1.6\times10^8 M_{\sun}$.

Colors, metallicities, and surface brightnesses of KH\,65,
KK\,180, and KK\,227 are similar to those of dSphs in
the Local Group, but the sizes are larger. Objects with similar
properties have recently been discovered in the Coma cluster
\citep{2015ApJ...798L..45V}. Ultra-diffuse galaxies (UDGs) are
large objects of extremely low density containing old stellar
populations. They are characterized by the following parameters
\citep{2015ApJ...798L..45V}: effective radii $R_{eff}=1.5 -
4.6$~kpc, absolute magnitudes $-16.0 \le M_g \le -12.5$, central
surface brightnesses $\mu_{g,0} = 24 - 26$~mag arcsec$^{-2}$,
masses $1\times10^7 - 3\times10^8 M_{\sun}${ }, and colors
$\langle g-i \rangle = 0.8 \pm 0.1$. The photometric data, masses,
and metallicities of KKH\,65 and KK\,227 are in an excellent
agreement with these values. Therefore, these two dSphs may be
classified as UDGs. KK\,180 is slightly brighter in the center
than typical UDGs.

Our spectroscopic study and grouping analysis allow us to measure
radial velocities of KKH\,65, KK\,180, and KK\,227 and derive
their group membership distances. We conclude that these three
galaxies are non-isolated.

It is worth noting that seven of nine candidates for isolated
dSphs from the list of \citet{2010Ap.....53..462K} have been
investigated up to now, including our three sample objects.
\cite{2014ApJ...782....4K} found out that KK258 is a very isolated
transitional-type LSB dwarf galaxy. I.D. Karachentsev
measured radial velocity of KKH9 and
confirmed it as a probable isolated dSph (private communication). KKR8 was not resolved
into stars on the HST images. It is much more distant than has
been thought before (L.N. Makarova, private communication, 2015).
KKR9 is a Galactic cirrus. Therefore, only two of the seven studied objects are isolated galaxies. 
In this work we did not expand the list of known isolated dSphs.
Objects of this class are extremely rare. 
More observational efforts are needed to establish the exact frequency
of their occurrence in the Local Supercluster.

\section*{Acknowledgments}
This work was performed with the support of the Russian Science
Foundation grant No 14-12-00965. We thank
Dodonov S.N. for the technical support of our observations. We
acknowledge the usage of the HyperLeda database
(http://leda.univ-lyon1.fr). Funding for the SDSS and SDSS-II was
provided by the Alfred P. Sloan Foundation, the Participating
institutions, the National Science Foundation, the U.S. Department
of Energy, the National Aeronautics and Space Administration, the
Japanese Monbukagakusho, the Max Planck Society, and the Higher
Education Funding Council for England. The SDSS Web Site is
http://www.sdss.org.

\bibliographystyle{aa}
\bibliography{26947_ap2c}
\clearpage
\appendix
\onecolumn
\section{SDSS surface photometry of KKH65, KK180, and KK227}
\label{SecA1} The data reduction\footnote{{\Large \bf
ELECTRONIC APPENDICES }} and analysis were conducted
using the \textsc{MIDAS} package. First, all images were cleaned
of all foreground and background objects.
Then we used the \textsc{SURFPHOT} program from the \textsc{MIDAS}
package to perform all the steps of the photometric procedure: the
sky background subtraction, the subsequent ellipse fitting and
integration of the light in the obtained ellipses. The algorithm
of this software is based on the formulas of
\citet{1987MitAG..70..226B} and \citet{1987AA...177...71B}. The
\textsc{FIT/FLAT\_SKY} task was used to approximate the sky
background by a surface created with a two-dimensional polynomial
and the least-squares method. All pixels of the residual image
with values that differed by more than two sigma from the mean
value of the sky were not used in the calculation of the
background level with the \textsc{FIT/BACKGROUND} program. The
typical accuracy of the sky brightness estimation was better than
1\,\%. This corresponds to the level $\textrm{SB}
\sim27\textrm{--}28$\,mag\,arcsec$^{-2}$ in the B band. The
ellipse fitting was carried out using the sky-subtracted images
and the task \textsc{FIT/ELL3}.
The \textsc{INTEGRATE/ELLIPSE} program was used to integrate the light
in the successive ellipses.
To transform SDSS magnitudes into the Johnson-Cousins system,
we used the empirical color
transformations by \cite{2006AA...460..339J}.
\begin{figure*}[h]
\begin{tabular}{p{0.3\textwidth}p{0.305\textwidth}p{0.3\textwidth}}
\includegraphics[width=0.23\textwidth,angle=0,clip]{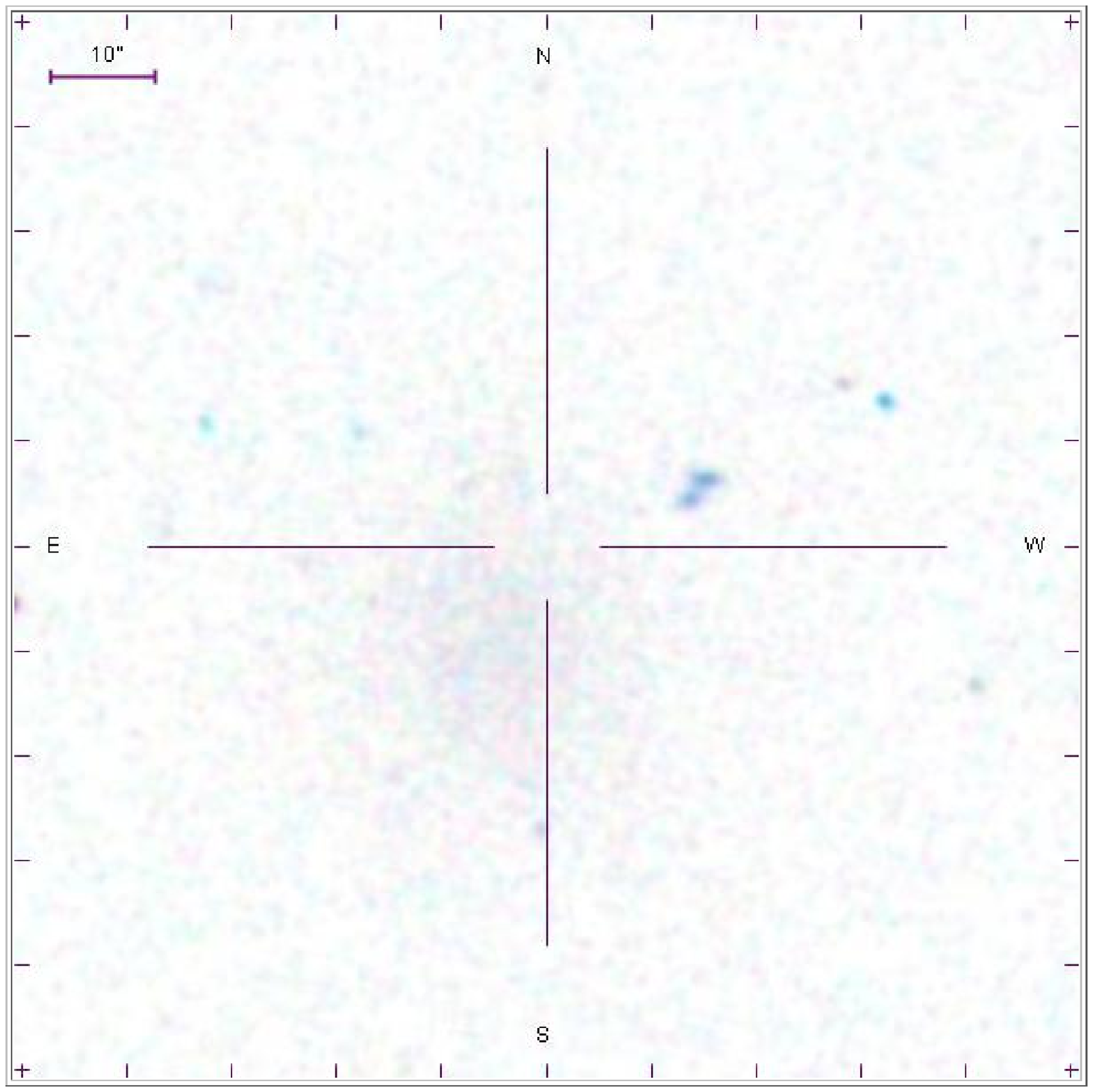}&
\includegraphics[width=0.235\textwidth,angle=0]{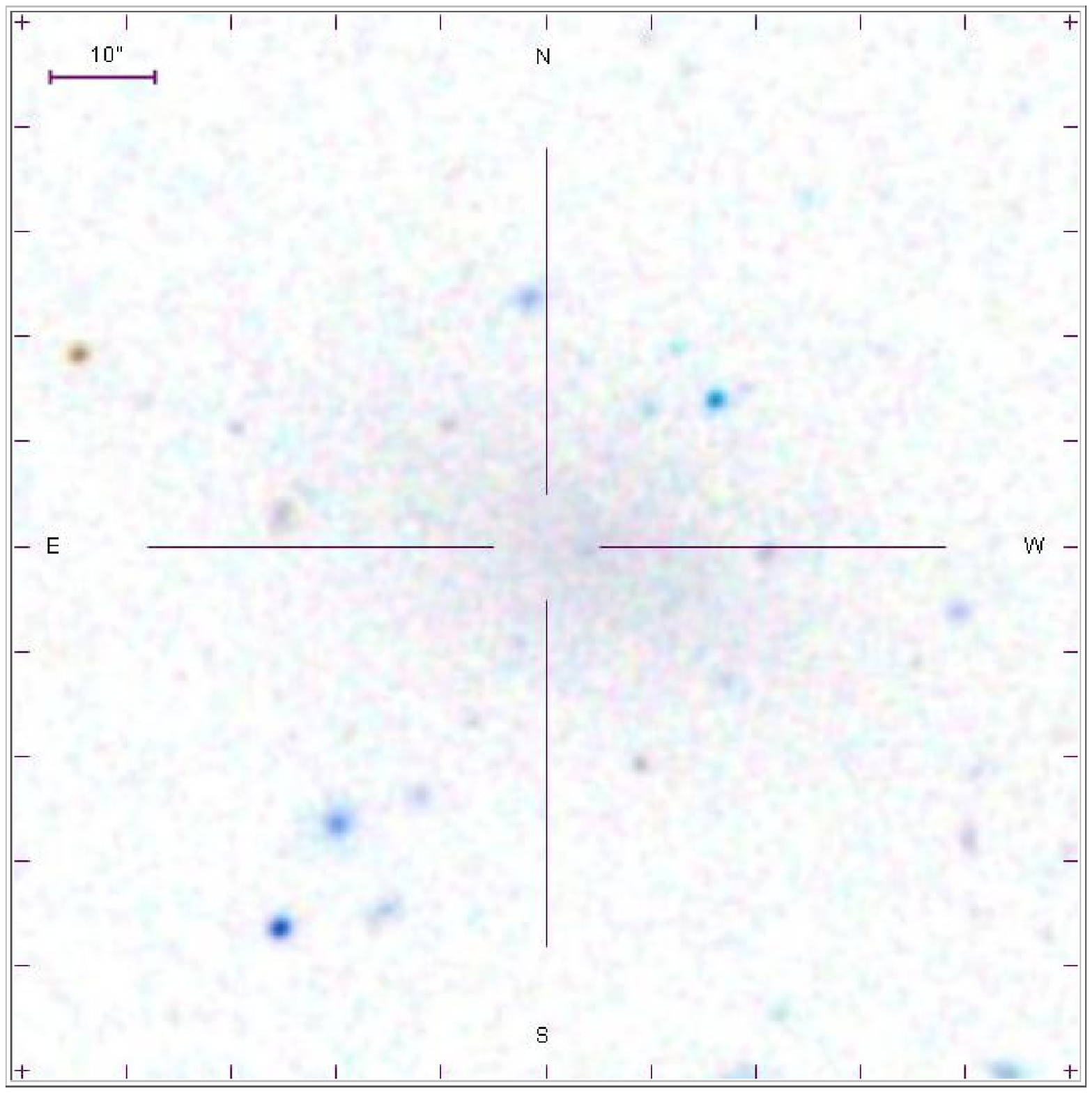}  &
\includegraphics[width=0.23\textwidth,angle=0]{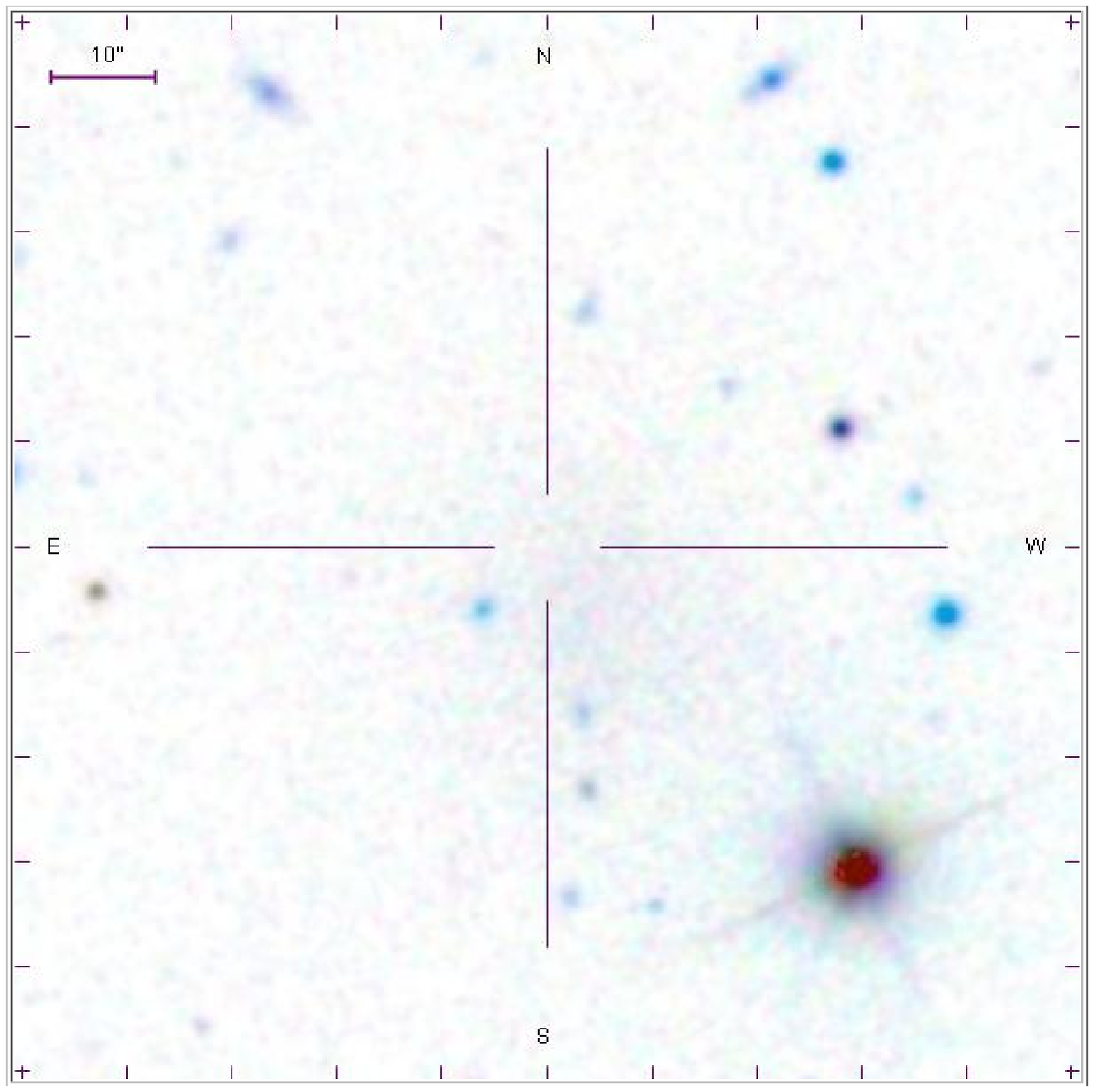}    \\
\end{tabular}
\caption{\label{fig:figures} SDSS images of the three dSphs KKH\,65, KK\,180, and KK\,227 (from left to right).}
\end{figure*}
\begin{table*}[h]
\caption{\label{tab_gri} Surface photometry using SDSS images in the g, r, i bands and Sersic function fitting results for
KKH~65, KK~180, and KK~227. The columns represent the following data:
(1) galaxy name, (2) limiting radius of our photometry in arcmin and axes ratio, (3) integrated magnitude within the limiting diameter,
(4) total Sersic model magnitude,
(5) integrated magnitude within the isophote 25 mag~arcsec$^{-2}$,
(6) mean surface brightness within the isophote 25 mag~arcsec$^{-2}$,
(7) Sersic index $n$ of the major-axis profile fit,
(8-10) corresponding Sersic model central surface brightness, effective radius in arcsec, and effective surface brightness,
and (9) integrated magnitude within the effective radius.}
\begin{tabular}{lrrrrrrrrr}\hline\hline
Name   & $R_{lim}$  & $g_l$  & $g_t$   &    $g_{25}$& $n_{g_s}$    &$\mu_{g,0_s}$ & $r_{e,g_s}$ &   $\mu_{e,g_s}$ &  $g_e$  \\
       & $b/a$     & $r_l$  & $r_t$   &    $r_{25}$& $n_{r_s}$    &$\mu_{r,0_s}$ & $r_{e,r_s}$ &   $\mu_{e,r_s}$ &  $r_e$  \\
       &            & $i_l$  & $i_t$   &    $i_{25}$& $n_{i_s}$    &$\mu_{i,0_s}$ & $r_{e,i_s}$ &   $\mu_{e,i_s}$ &  $i_e$  \\

 (1)   &  (2)     & (3)   & (4)   &    (5)     & (7)          &  (8)         & (9)         &  (10)           &  (11)      \\ \hline
KKH\,65  &  0.67  & 17.14   & 17.05   & 19.60  & 0.88$\pm$0.03 & 24.69$\pm$0.02 & 20.52$\pm$0.42  & 26.25       & 17.79  \\
         &  0.70  & 16.83   & 16.82   & 17.88  & 0.77$\pm$0.02 & 24.11$\pm$0.02 & 16.22$\pm$0.31  & 25.45       & 17.58  \\
         &        & 16.57   & 16.43   & 17.19  & 0.85$\pm$0.02 & 23.72$\pm$0.02 & 17.21$\pm$0.35  &  25.23      & 17.15  \\
         &       &          &         &         &              &                &                 &             &        \\
KK\,180 &  0.75 & 16.52   & 16.42   &  17.89  &  0.97$\pm$0.02 & 23.96$\pm$0.02 & 21.06$\pm$0.29  &  25.71      & 17.13  \\
        & 0.71  & 16.02   & 15.89   &  16.66  &  0.98$\pm$0.02 & 23.36$\pm$0.02 & 20.58$\pm$0.35  &  25.15      & 16.63  \\
        &       & 15.58   & 15.52   &  16.10 &   0.94$\pm$0.02 & 23.14$\pm$0.02 & 21.27$\pm$0.40  &  24.82      & 16.23  \\
        &       &         &        &         &                 &                &                 &             &        \\
KK\,227 & 0.48 & 17.49   & 17.22   &  --     &   0.76$\pm$0.03  & 24.99$\pm$0.02 & 20.00$\pm$0.50  &  26.29     &  18.00 \\
        & 0.81 & 17.10   & 16.85   &  18.05  &   0.82$\pm$0.03  & 24.24$\pm$0.03 & 17.51$\pm$0.51  &  25.66      & 17.57  \\
        &      & 16.62   & 16.34   &  17.35  &   0.84$\pm$0.03  & 23.88$\pm$0.02 & 19.15$\pm$0.54  &  25.37      & 17.12  \\
 \hline\hline

\end{tabular}
\end{table*}
\newpage
\begin{figure*}
\includegraphics[width=0.3\textwidth,angle=-90]{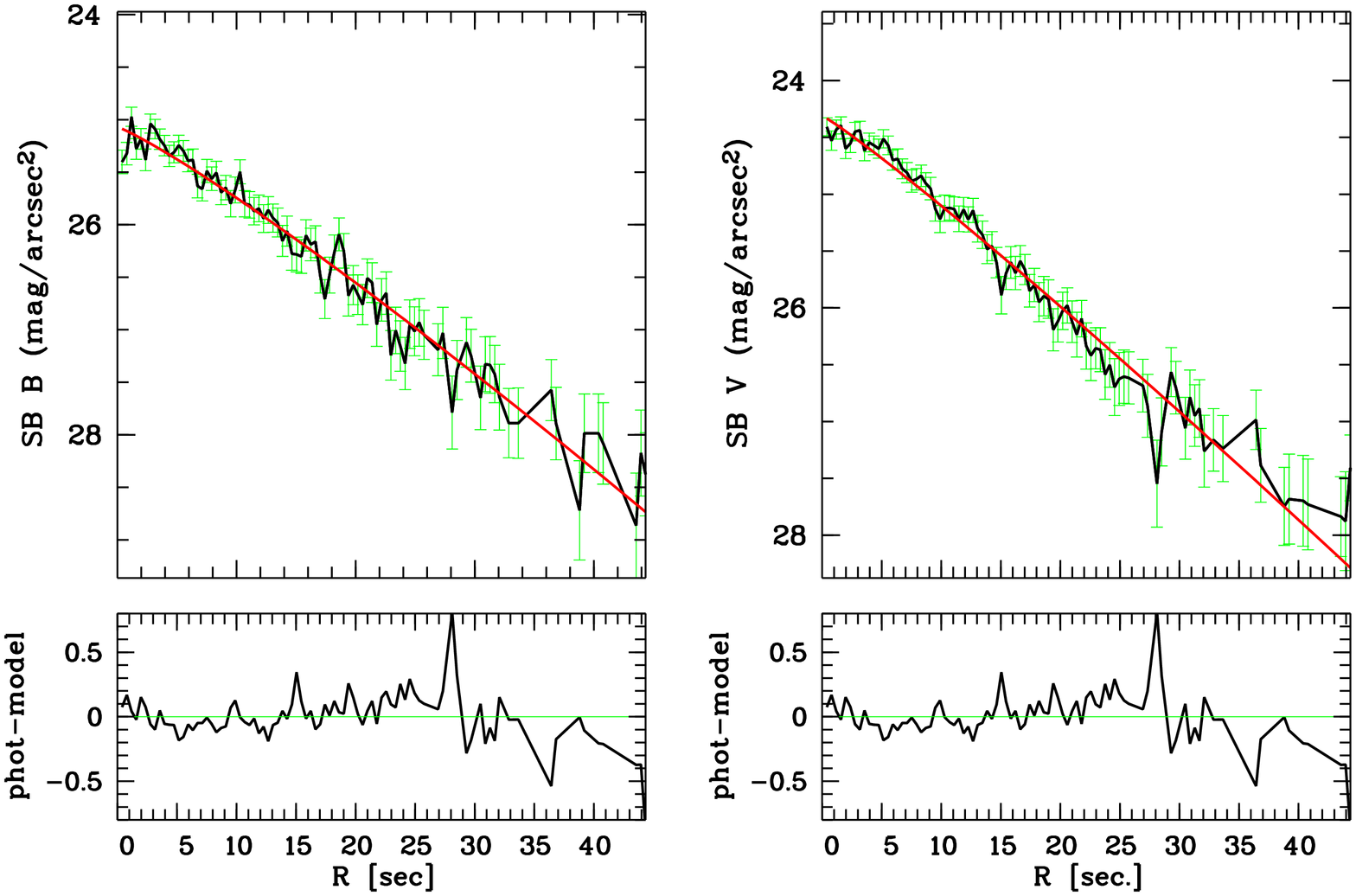}
\includegraphics[width=0.3\textwidth,angle=-90]{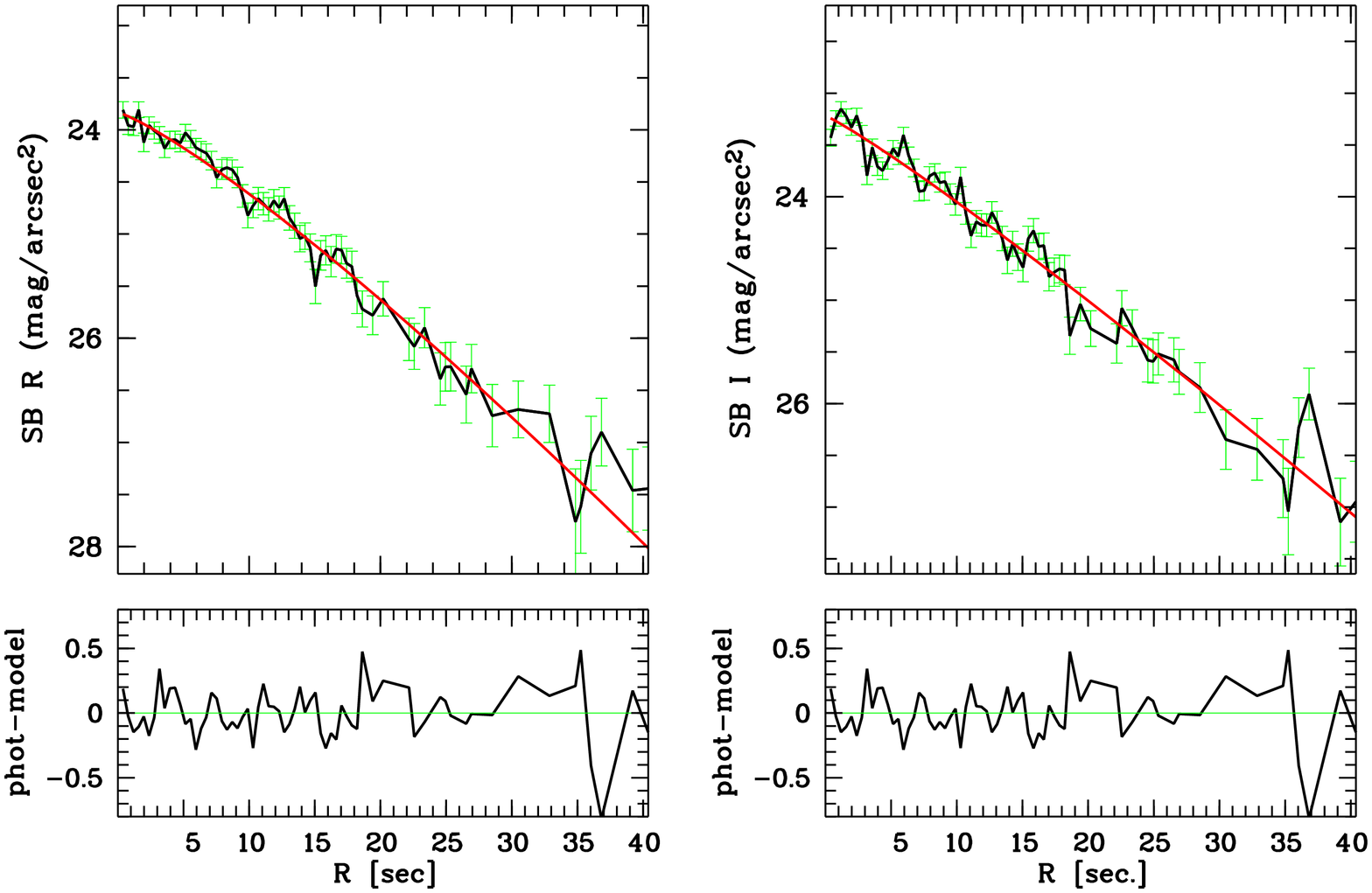}
\caption{\label{fig:serskkh65} Sersic function fits to the equivalent profiles
of KKH\,65 in the $B$, $V$, $R$, $I$ bands.}
\end{figure*}
\begin{figure*}
\includegraphics[width=0.3\textwidth,angle=-90]{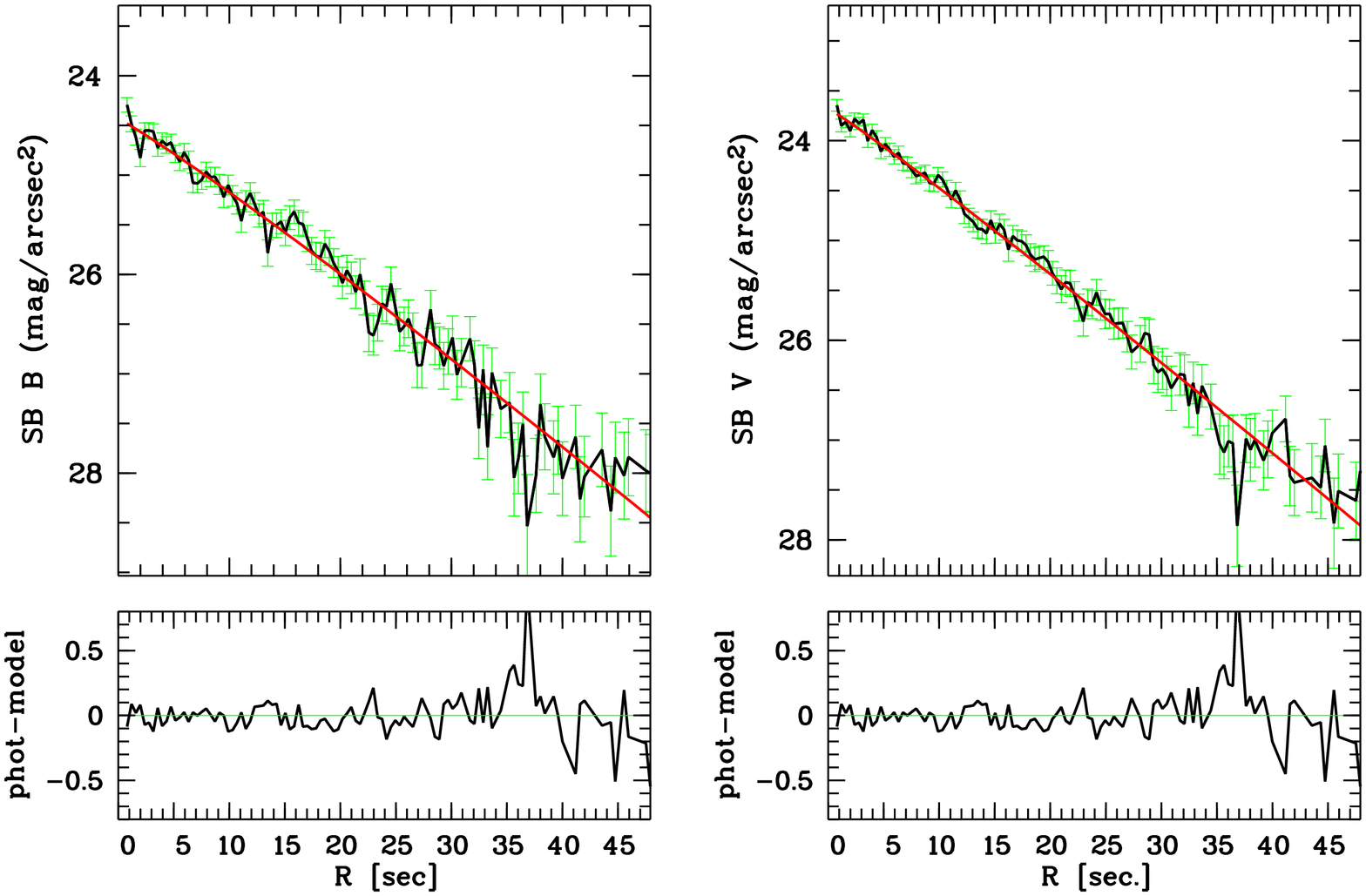}
\includegraphics[width=0.3\textwidth,angle=-90]{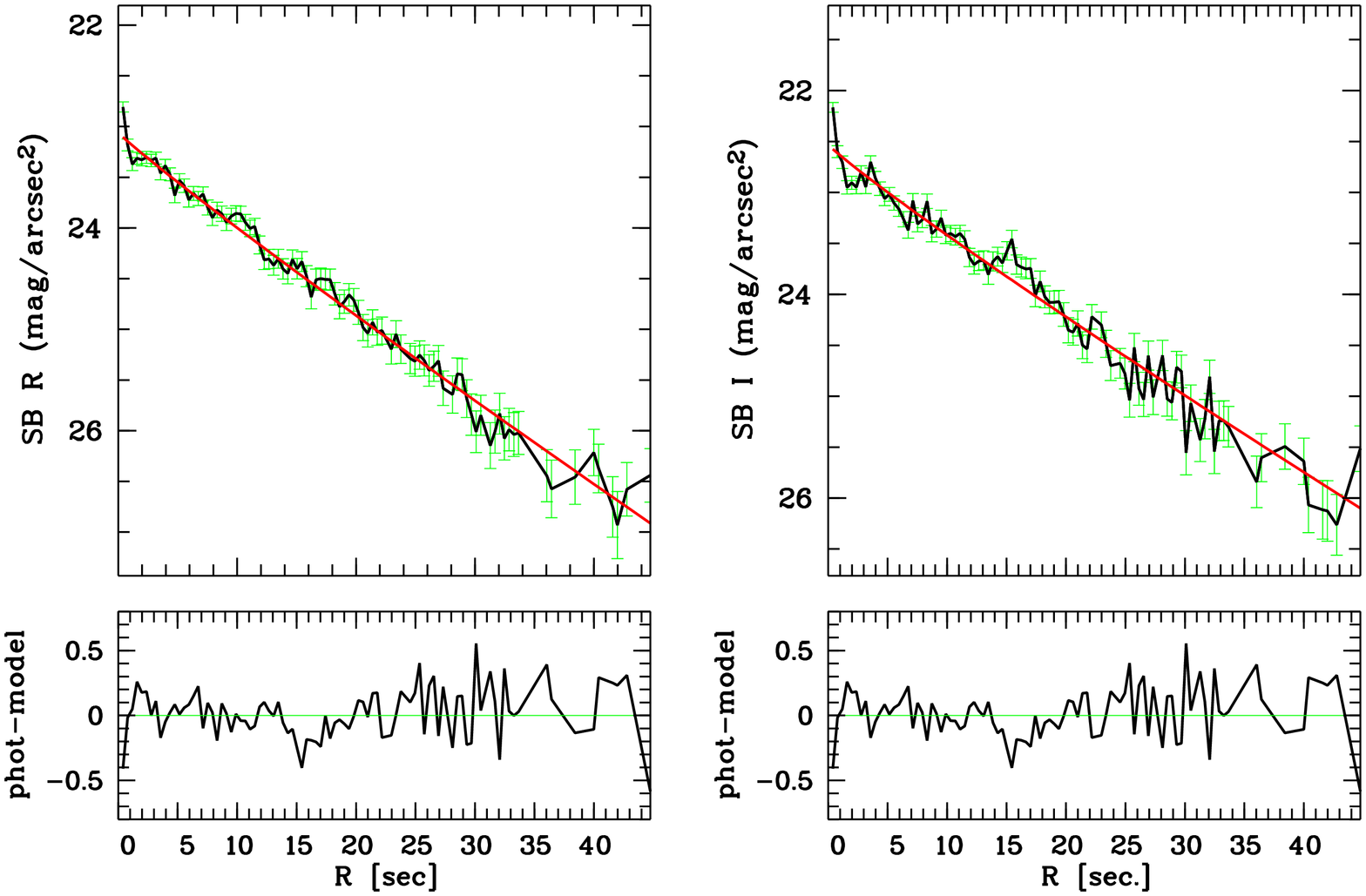}
\caption{\label{fig:serskk180} Same as Fig. A.2,
but for KK\,180. }
\end{figure*}
\begin{figure*}
\includegraphics[width=0.3\textwidth,angle=-90]{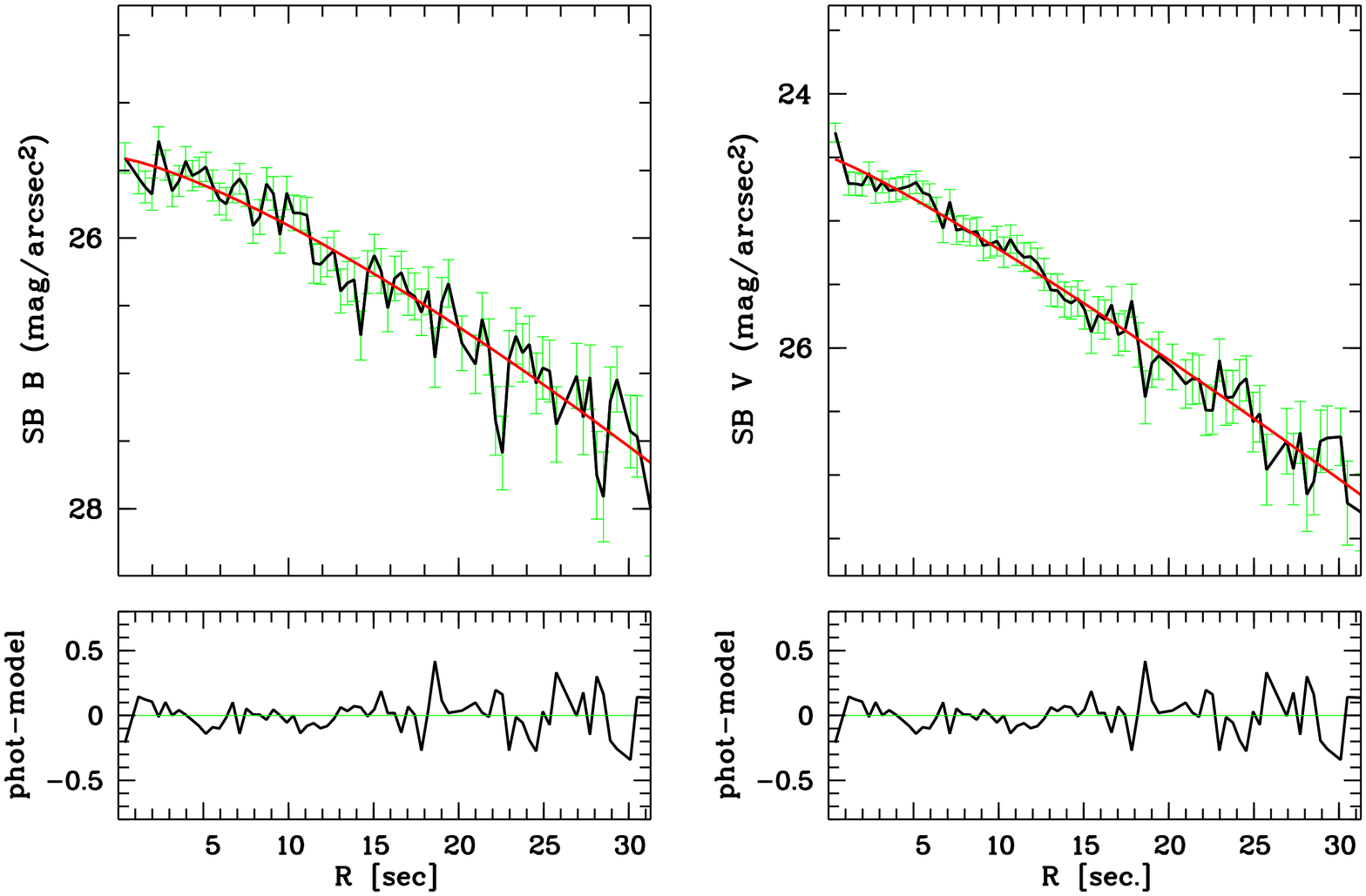}
\includegraphics[width=0.3\textwidth,angle=-90]{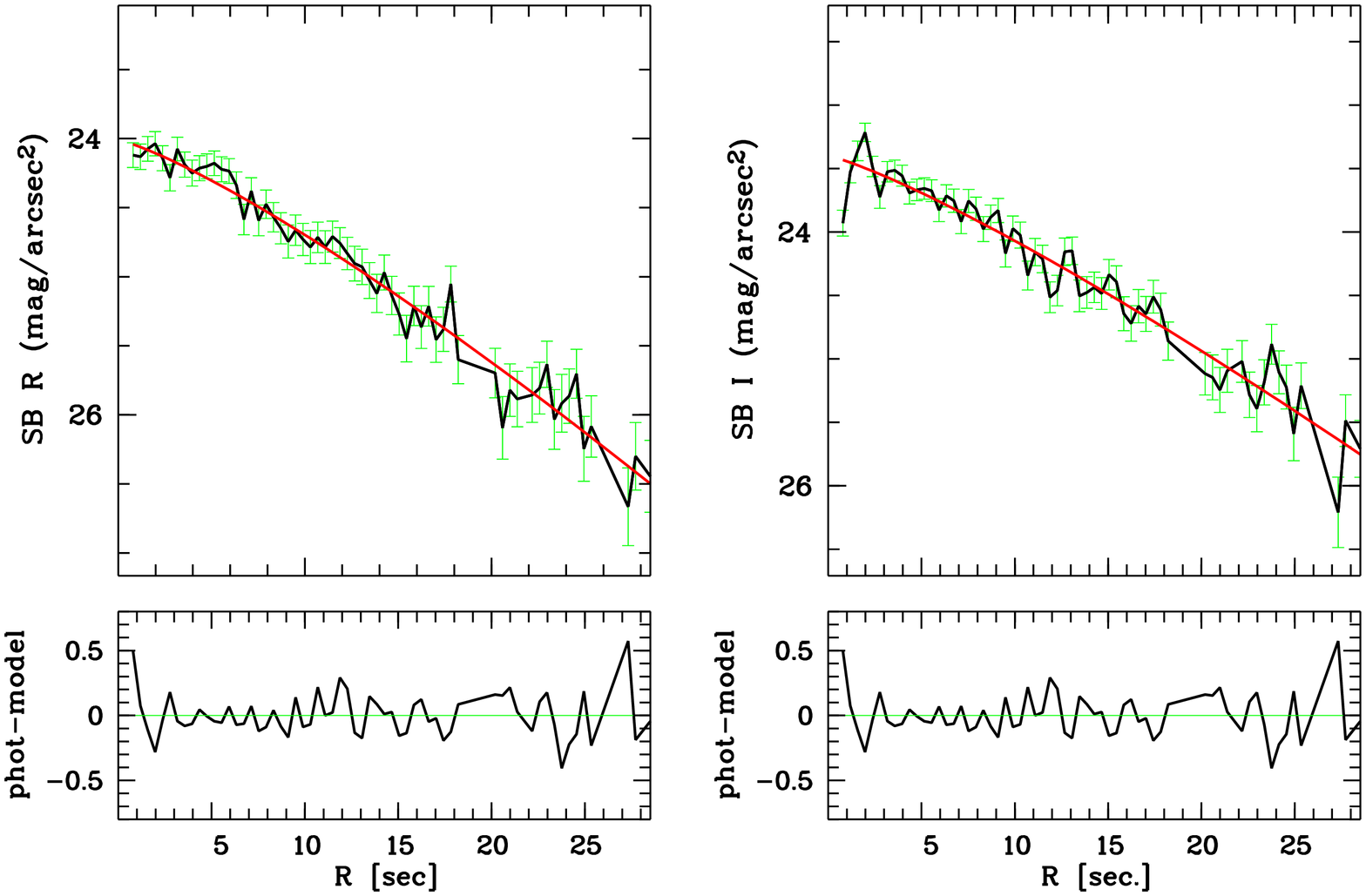}
\caption{\label{fig:serskk227} Same as Fig. A.3,
but for KK\,227.}
\end{figure*}

\clearpage
\section{Members of the groups containing KKH\,65 and KK180}
\label{association}
\subsection{NGC\,3414 group}
The algorithm developed by MK11 identified eight members of the NGC\,3414 group. Using
the same group-finding algorithm and new observational data
(radial velocities from SDSS DR9 and spectroscopic and photometric
data for KKH\,65), we found 19 galaxies with differences in radial
velocities $\vert V^{N3414}_{LG} - V^{gal}_{LG}\vert$ smaller than
600~km~$s^{-1}$ and within the projected radius from NGC3414
$\textrm{R}^{proj}_{N3414} = 600$~kpc (Table~\ref{tab:n3414gr}).
Eleven of them are members of the group with $ii< 1,$
and eight objects are candidate field galaxies with $ii> 1$. Some
of $log(ii)$ values are close to zero. This means that the actual
group size may change if accurate redshift-independent distances
will be available. After selecting the group members, we compared
the properties of KKH\,65 and the surrounding galaxies. There are
three candidate early-type group member galaxies fainter than
KKH\,65, and six candidate field early-type objects comparable to
KKH\,65 (Table~\ref{tab:n3414gr}). SDSS images and our photometric
study indicate that KKH\,65 is the most diffuse among all these
objects.

Table~\ref{tab:n3414gr} lists the data for the projected neighbors of KKH\,65
with the differences in radial velocities between NGC~3414 and each individual galaxy that are smaller than $600$~km$s^{-1}$:
(1) galaxy name, (2) equatorial coordinates,
(3) morphological type in numerical code according to the RC3 catalog \citep{1991rc3..book.....D},
(4) integrated magnitude in the $Ks$ band, corrected for Galactic extinction,
(5) radial velocity with respect to the centroid of the LG,
(6) projected separation in kpc from NGC3414 assuming that the galaxies are located at their Hubble distances,
(7) logarithmic isolation index calculated with respect to NGC\,3414 calculated according to MK11.
\begin{table*}[h]
\caption{\label{tab:n3414gr} S0 galaxy NGC3414 and galaxies with differences in radial velocities
$\vert V^{N3414}_{LG} - V^{gal}_{LG}\vert$ smaller than 600~km~$s^{-1}$ and within the projected radius
from NGC3414 $\textrm{R}^{proj}_{N3414} = 600$~kpc.}
\scriptsize
\begin{tabular}{lcrrrcc} \hline\hline
Galaxy & RA (J2000.0) DEC  & T & $K_s{_0}$ & $V_{LG}$ & $\textrm{R}^{proj}_{N3414}$ & $ log(ii)$ \\
 &  \hspace{0.12cm} $h$  \hspace{0.08cm} $m$  \hspace{0.05cm} $s$ \hspace{0.6cm} $ ^{o}$ \hspace{0.1cm} $ '$ \hspace{0.1cm} $''$ &  & (mag) & (km/s) & (kpc) & (N3414) \\
\hline
\smallskip
 NGC3414                  & 10 51 16.21 +27 58 30.4 &  -2  &  8.03 & 1351 &   0 &         \\ 
UGC5844                   & 10 43 55.98 +28 08 50.1 &  7   & 13.9  & 1398 & 535 & -0.32 \\ 
UGC5921                   & 10 49 12.79 +27 55 21.3 &  8   & 13.48 & 1356 & 148 & -2.23 \\ 
NGC3400                   & 10 50 45.47 +28 28 08.6 &  1   & 10.20 & 1344 & 163 & -2.16 \\ 
PGC4576692                & 10 50 49.39 +28 03 30.6 &  9   & 16.5  & 1460 &  43 & -0.69 \\ 
PGC1823017                & 10 51 07.88 +27 58 46.5 &  -1  & 18.0  & 1779 &  12 & -0.06 \\ 
$[KK90]$013, PGC93597     & 10 51 13.01 +28 00 22.3 &  -3  & 12.38 & 1125 &  10 & -0.66 \\ 
UGC5958                   & 10 51 15.87 +27 50 55.6 &  4   & 12.4  & 1113 &  37 & -0.04 \\ 
SDSS J105123.21+280125.7  & 10 51 23.21 +28 01 25.7 &  -3  & 15.8  & 1646 &  20 & -0.15 \\ 
NGC3418                   & 10 51 23.95 +28 06 43.2 &   0  & 10.38 & 1187 &  42 & -0.34 \\ 
BTS22                     & 10 51 37.14 +27 49 19.6 &  -3  & 14.4  & 1256 &  53 & -0.67 \\ 
\smallskip
{\bf KKH\,65$=$BTS23}      & 10 51 59.00 +28 21 45.0 &  -3  & 15.3  & 1282 & 132 & -0.58 \\ 
PGC4251293                & 10 46 30.44 +27 41 56.2 &  10  & 15.5  & 1468 & 367 &  0.31 \\ 
NGC3380                   & 10 48 12.17 +28 36 06.4 &  1   &  9.99 & 1528 & 317 &  0.53 \\ 
PGC4558491                & 10 50 13.88 +27 18 03.9 & -3   & 16.0  & 1604 & 252 &  0.82 \\ 
BTS24                     & 10 52 00.35 +27 45 32.7 &  -2  & 15.6  & 1591 & 95  &  0.29 \\ 
PGC1813510                & 10 52 14.12 +27 37 23.8 &  -1  & 15.2  & 1511 & 141 &  0.16 \\ 
SDSS J105218.32+272314.2  & 10 52 18.32 +27 23 14.2 &  10  & 15.9  & 1595 & 222 &  0.79 \\ 
NGC 3451                  & 10 54 20.97 +27 14 25.1 &  7   & 10.2  & 1259 & 313 &  0.04 \\ 
PGC4572094                & 10 54 41.75 +28 07 31.1 &  9   & 15.9  & 1537 & 266 &  0.57 \\ 
\hline\hline
\end{tabular}
\end{table*}
\subsection{UGC8036 group}
Table~\ref{tab:kk180gr} lists galaxies located within the projected vicinity from KK\,180.
The results of the MK11 algorithm (Table~\ref{tab:kk180gr}) favor an extremely weak gravitational influence between the UGC\,8036 group members.
At the same time, the gravitational attraction from the Virgo cluster is strong.
(1) Galaxy name, (2) equatorial coordinates, (3) morphological
type in numerical code according to de Vaucouleurs et al. (1991),
(4) integrated magnitude in the $Ks$ band, (5) radial velocity
with respect to the centroid of the LG, (8) projected separation
in kpc from UGC\,8036, (9) isolation index with respect to
UGC\,8036, (10) projected separation in Mpc from M87, (11)
isolation index calculated with respect to the center of the Virgo
cluster with the total mass of $8\times 10^{14} M_{\odot}$
\citep{2014ApJ...782....4K}. The projected separations are
calculated assuming that the galaxies are at their Hubble
distances.
\begin{table*}[b]
\caption{\label{tab:kk180gr} Most massive galaxy in the Virgo cluster M87 and six galaxies around
KK180 with a radial velocity difference smaller than $\vert V^{M87}_{LG} - V^{gal}_{LG}\vert <600$~km~$s^{-1}$
and within the projected radius from KK180 $\textrm{R}^{proj}_{KK180} = 600$~kpc.}
\scriptsize
\begin{tabular}{lcrrrcccc}\hline\hline
Galaxy                    & RA (J2000.0) DEC       & T& $Ks{_0}$ & $V_{LG}$& $\textrm{R}^{proj}_{U8036}$& $log(ii)$& $\textrm{R}^{proj}_{VC}$   & $log(ii)$ \\
 &  \hspace{0.11cm} $h$  \hspace{0.08cm} $m$  \hspace{0.05cm} $s$ \hspace{0.5cm} $ ^{o}$ \hspace{0.1cm} $ '$ \hspace{0.1cm} $''$ &  & mag & km/s& Mpc & U8036&  Mpc &  VC \\
\hline
                           &                         &    &        &      &        &     &       &       \\
 NGC4486=M87               & 12 30 49.41 +12 23 28.5 & -4 &  5.4   & 1165 &   --   & --  & 0.00  & --    \\
          UGC8036          & 12 54 48.60 +19 10 37.6 &  6 & 11.96  &  844 &   0.00 & --  & 1.27  & -1.57 \\
 SDSS J125651.47+163024.2  & 12 56 51.50 +16 30 24.2 & -2 & 13.39  & 1157 &   0.60 & 3.30& 1.07  & -1.26 \\
$[KK98]$ 173,  LSBC D575-07& 12 58 35.97 +17 48 46.9 & 9  & 18.70  &  934 &   0.35 & 2.08& 1.23  & -1.83 \\
 SDSS J130320.36+175909.7  & 13 03 20.43 +17 59 09.1 & -2 & 15.28  &  926 &   0.50 & 2.13& 1.38  & -1.75 \\
{\bf KK\,180}              & 13 04 30.00 +17 45 32.0 & -3 & 14.14  &  609 &   0.47 & 3.01& 1.39  & -1.20 \\
  SDSS J130440.05+184438.7 & 13 04 40.05 +18 44 39.1 & 10 & 15.71  &  766 &   0.46 & 2.06& 1.48  & -1.36 \\
\hline\hline
\end{tabular}
\end{table*}

\clearpage
\section{QSO projected on KK\,227}
\label{sec_QSO}
\begin{figure*}
\begin{tabular}{p{0.5\textwidth}p{0.5\textwidth}}
\includegraphics[width=0.43\textwidth,angle=180]{QSO.ps} &
\hspace{-1.7cm}
\includegraphics[width=0.4\textwidth,angle=-90]{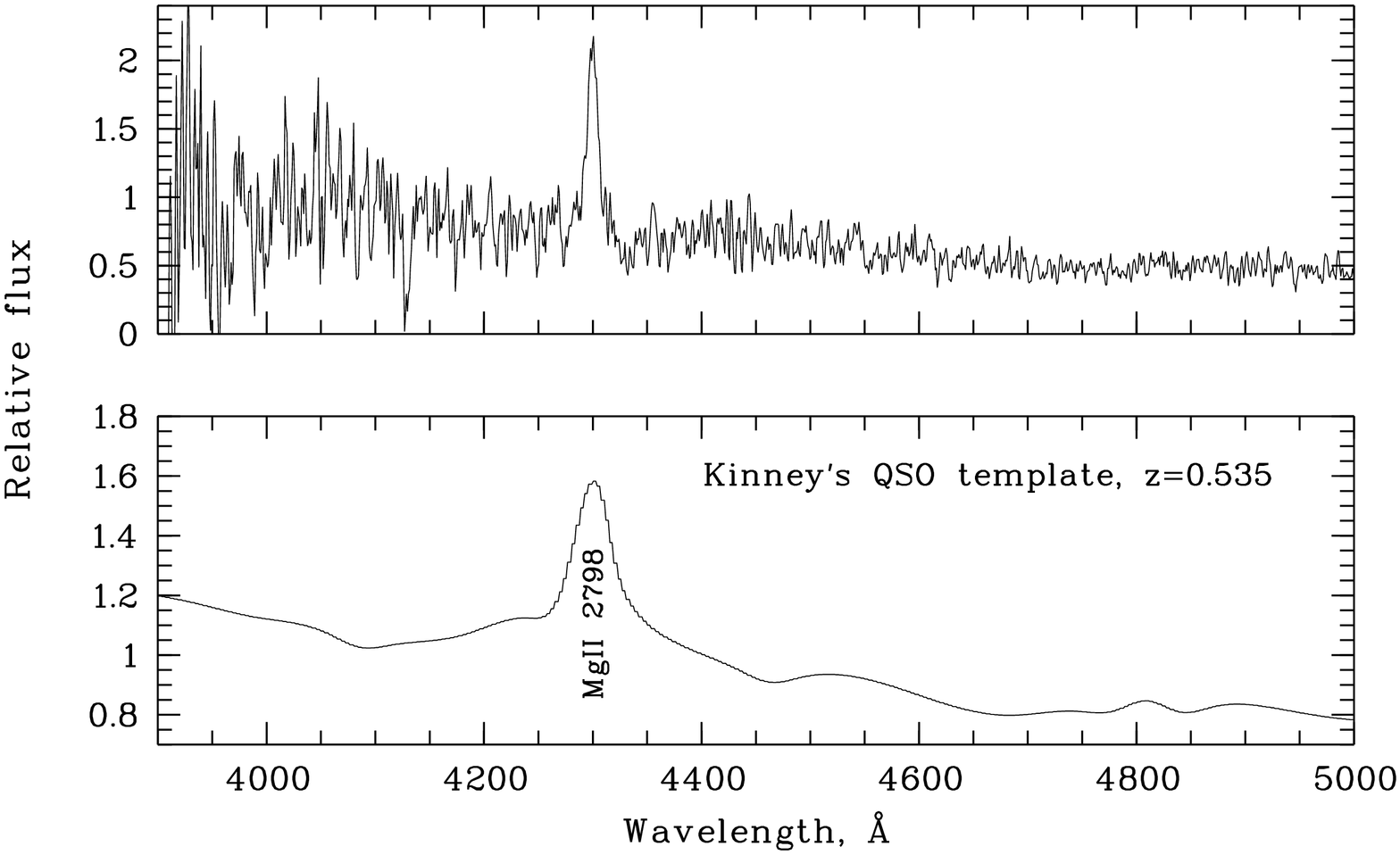} \\
\vspace{-3cm}
\end{tabular}
\caption{\label{fig:QSO}Left: Settings of the slit on the
preliminary short-exposure image of KK\,227 in the $B$ band. The
approximate location of a quasar is circled. Right:
One-dimensional total spectrum of the quasar after observations in
the two positions of the slit.}
\end{figure*}

 The long-slit observations
revealed a distant extended object seen through
the stellar body of KK\,227 with a QSO-like spectrum. It was
detected in both slit positions. We estimated its redshift
$z=0.535$ using only one clearly seen line MgII 2798\AA\ . The
orientation of the slit in the case of KK\,227 and the location of
the quasar are shown in Fig.~\ref{fig:QSO}.
\end{document}